# Repetitive single electron spin readout in silicon

J. Yoneda[1,2]*, K. Takeda[1], A. Noiri[1], T. Nakajima[1], S. Li[1], J. Kamioka[3], T. Kodera[3], S. Tarucha[1]*

[1]*RIKEN Center for Emergent Matter Science, RIKEN, Saitama, 351-0198 Japan.*

[2]*Center for Quantum Computation and Communication Technology, School of Electrical Engineering and Telecommunications, The University of New South Wales, Sydney, NSW 2052, Australia*

[3]*Department of Electrical and Electronic Engineering, Tokyo Institute of Technology, Tokyo, 152-8550 Japan.*

*Correspondence to: jun.yoneda@alum.riken.jp, tarucha@riken.jp

**Single electron spins confined in silicon quantum dots hold great promise as a quantum computing architecture with demonstrations of long coherence times[1], high-fidelity quantum logic gates[2-4], basic quantum algorithms[5] and device scalability[6]. While single-shot spin detection is now a laboratory routine[1-7], the need for quantum error correction in a large-scale quantum computing device demands a quantum non-demolition (QND) implementation[8-10]. Unlike conventional counterparts, the QND spin readout imposes minimal disturbance to the probed spin polarization and can therefore be repeated to extinguish measurement errors. However, it has remained elusive for an electron spin in silicon as it involves exquisite exposure of the system to the external circuitry for readout while maintaining the coherence and integrity of the qubit. Here we show that an electron spin qubit in silicon can be measured in a highly non-demolition manner by probing another electron spin in a neighboring dot Ising-coupled to the qubit spin. The high non-demolition fidelity (99% on average) enables over 20 readout repetitions of a single spin state, yielding an overall average measurement fidelity of up to 95% within 1.2 ms. We further demonstrate that our repetitive QND readout protocol can realize heralded high-fidelity (> 99.6%) ground-state preparation. Our QND-based measurement and preparation, mediated by a second qubit of the same kind, will allow for a new class of quantum information protocols with electron spins in silicon without compromising the architectural homogeneity.**

The ability to measure a quantum system in a single-shot QND manner has a pivotal role in quantum error correction and quantum information processing, as well as being central to quantum science[8-10]. An ideal single-shot QND readout process would, in addition to yielding an eigenvalue of the observable with projection probability for the input state (measurement), leave the system in the projected input state (non-demolition), meaning that the measurement is repeatable and that a posterior state can be predicted based on the eigenvalue obtained (preparation)[8]. These features contrast with conventional readout schemes of a silicon spin qubit, which inherently demolish the spin state by mapping it to a more readily detectable, charge degree of freedom[1-7]. Such spin-to-charge conversion mechanisms are employed to facilitate to measure the small magnetic moment of a single electron spin within its relaxation time, which, although exceptionally long for a solid-state quantum system, is limited to the millisecond timescale. Synthesizing an ancilla system which can be repeatedly initialized, controlled conditionally on the qubit state and separately measured, all on the microsecond timescale, constitutes a major challenge for the QND readout of a silicon electron spin qubit.

In this work we demonstrate repeatable measurements of a silicon electron spin qubit. We use a neighboring electron spin as an ancilla, with which we can perform a QND qubit readout at a 60 μs repetition cycle through a conditional rotation and spin-selective tunneling. The highly QND nature is evidenced by the strong correlation between successive ancilla measurement outcomes. We take advantage of the repeatability and construct a QND qubit readout from $n$ consecutive ancilla measurements to improve the overall performance. For complete characterization as a QND readout process, we identify and evaluate three key metrics[8]: the non-demolition fidelity ($F_{QND}$ = 99% for $n$ = 1); the measurement fidelity ($F_M$ = 95% for $n$ = 20); the preparation fidelity ($F_P$ = 92% for $n$ = 20). (The numbers are the average of the spin-down and -up cases.) The non-demolition and preparation fidelities ($F_{QND}$ and $F_P$) which are dissimilar to those in the destructive readout illustrate the distinct properties of the QND readout. We further show that the repetitive readout scheme allows us to preselect the cases where the qubit state is prepared with fidelities > 99.6%.

Our qubit and ancilla are electron spins confined in a double Si/SiGe quantum dot (Fig. 1a) with natural isotopic abundance[11]. Spin states can be discriminated and reinitialized within 30 μs relying on energy-selective spin-to-charge conversion and the reflectometry response from a neighboring charge sensor[7,11]. An on-chip micromagnet magnetized in an external magnetic field $B_{ext}$ = 0.51 T separates the resonance frequencies of the qubit and ancilla spins by 640 MHz (centered around ~16.3 GHz). This enables frequency-selective electric-dipole-spin resonance rotations of individual spins at several MHz and ensures that the exchange interaction of ~ MHz is well represented by the Ising type with minimal disturbance to the spin polarizations[12,13].

We correlate the ancilla and the qubit spins by a controlled-rotation gate (Fig. 1b). During a square gate-voltage pulse for a duration $t_{CZ}$ at a symmetric operation point, the ancilla spin acquires a qubit-state-dependent phase due to enhanced exchange coupling[3,14]. A Hahn echo sequence converts this phase to the ancilla spin polarization, in a robust manner against a slow drift of the ancilla precession frequency and the qubit-state-independent phase induced by the square gate-voltage pulse (~20π per μs) and the microwave bursts (~0.16π)[15,16]. We extract the qubit-dependent phase shift by changing the prepared qubit state by the microwave burst time $t_b$ (Fig. 1c). The extracted phase grows linearly with $t_{CZ}$ (Fig. 1d), consistent with an induced excess exchange coupling $J$ of 0.94 MHz. Choosing $t_{CZ}$ = 0.53 μs and an appropriate projection phase $\theta$, we can implement a conditional rotation which maps the qubit state to the ancilla spin, allowing for the ancilla-based measurement of the qubit spin.

We now demonstrate that the ancilla can be repeatedly entangled with the qubit and measured, using a sequence shown in Fig. 2a. After preparing the qubit state by microwave control, we repeat 30 cycles of a controlled-rotation gate and the ancilla measurement and reinitialization, until we destructively read out and reinitialize the qubit. We use $m_i$ and $q$ to denote the outcomes of the $i$-th ancilla measurement (with $i$ = 1, 2, ... 30) and the final qubit readout, respectively. Remarkably, all ancilla measurement outcomes show clear Rabi oscillations (Fig. 2b), indicating each functions as a single-shot QND readout of the qubit. Strong correlations between successive measurements, a hallmark of the QND readout, are verified from joint probabilities $P(m_1 m_2)$, see Fig. 2c.

The Rabi oscillation visibility of $m_i$ is affected by both the probability distribution $p_{i-1}^{\downarrow(\uparrow)}$ of the prepared qubit spin state $s_{i-1}$ and the $i$-th QND measurement fidelity $f_i^{\downarrow(\uparrow)}$ given $s_{i-1} = \downarrow (\uparrow)$. We separate these preparation and measurement errors[17] by expressing the joint probability $P(m_i m_{30})$ as

$$P(m_i m_{30}) = \sum_{s=\downarrow,\uparrow} p_{i-1}^s \Theta_{s,m_i}(f_i^s) \Theta_{s,m_{30}}(g_i^s). \quad (1)$$

Here $g_i^{\downarrow(\uparrow)}$ denotes the measurement fidelity of $m_{30}$ for $s_{i-1}$ prepared in $\downarrow (\uparrow)$, and $\Theta_{s,m}(f)$ equals $f$ when $s = m$ and $1 - f$ when $s \neq m$. We model $p_{i-1}^{\downarrow(\uparrow)}$ by an exponentially decaying Rabi oscillation and obtain $p_i^{\downarrow(\uparrow)}$, $f_i^{\downarrow(\uparrow)}$ and $g_i^{\downarrow(\uparrow)}$ as a function of $i$ (see Methods). We find that $f_i^{\downarrow(\uparrow)}$ is essentially $i$-independent as expected, with the average 85% (75%) for $i$ = 1-20.

A distinct feature of the QND readout is that it is repeatable, meaning we can potentially gain more accurate information about the qubit state from consecutive measurements. In the following, we leverage this potential by constructing a cumulative QND readout from $n$ outcomes, $\boldsymbol{m}_n = \{m_1, m_2, \ldots m_n\}$ which yields estimators σ for $s_0$ (the input qubit state, projected to either spin-down or -up) and ς for $s_n$ (the posterior qubit state), see Fig. 3a. We characterize its performance as a QND readout as a function of $n$, through three key fidelity figures of merit, $F_{QND}$, $F_M$ and $F_P$. These fidelities are, as depicted in Fig. 3a, defined by the correspondences between the estimators (σ and ς) and/or the qubit states before and after the process ($s_0$ and $s_n$). Importantly, these together will enable us to test all key criteria that the QND readout should satisfy[8] – i.e. non-demolition ($F_{QND}$), measurement ($F_M$) and preparation ($F_P$).

We first assess the non-demolition fidelity $F_{QND}^{\downarrow(\uparrow)}$, which addresses the requirement that the measured observable (spin-down or up) should not be disturbed. It represents the correlation between the projected input ($s_0$) and posterior ($s_n$) qubit states, and can be defined using the conditional probability of $s_n$ given $s_0$ as $F_{QND}^{\downarrow(\uparrow)} = P(s_n = s_0 | s_0 = \downarrow (\uparrow))$. It follows from this definition that $p_n^\downarrow = F_{QND}^\downarrow p_0^\downarrow + (1 - F_{QND}^\uparrow) p_0^\uparrow$. The results obtained from the fit to this equation is shown in Fig. 3b, where $F_{QND}^{\downarrow(\uparrow)}$ gradually decreases to 99% (61%) as $n$ is increased up to 20. By modeling the $n$ dependence of $p_n^\downarrow$ (see Methods), we estimate $F_{QND}^{\downarrow(\uparrow)}$ for $n$ = 1 to be 99.92% (97.7%), corresponding to the longitudinal spin relaxation time $T_1^{\downarrow(\uparrow)}$ of 78 ms (2.5 ms) given the 60 µs cycle time.

The second requirement for the QND readout is that the measurement result should be correlated with the input state following the Born rule. We test this through the measurement fidelity defined as $F_M^{\downarrow(\uparrow)} = P(\sigma = s_0 | s_0 = \downarrow (\uparrow))$, where $\sigma$ is the estimator for the input qubit state $s_0$ based on measurement results $\mathbf{m}_n$. When $\sigma$ is the more likely value of $s_0$, $P(\mathbf{m}_n | s_0 = \sigma) > P(\mathbf{m}_n | s_0 = \bar{\sigma})$ with $\bar{\sigma}$ denoting the spin opposite to $\sigma$. We calculate these likelihoods using a Bayes model that assumes spin-flipping events (see Methods). $\sigma$ shows larger Rabi oscillations as $n$ is increased (Fig. 3c), demonstrating $F_M^{\downarrow(\uparrow)}$ enhancement by repeating ancilla measurements in our protocol. We obtain $F_M^{\downarrow(\uparrow)}$ (Fig. 3d) through $P(\sigma = \downarrow) = F_M^\downarrow p_0^\downarrow + (1 - F_M^\uparrow) p_0^\uparrow$. While $F_M^{\downarrow(\uparrow)} =$ 88% (73%) for $n = 1$, it reaches 95.6% (94.6%) for $n = 20$, well above the measurement fidelity threshold for the surface code[9].

The last feature of the QND readout to be evaluated is the capability as a state preparation device. In order to quantify how precisely our cumulative QND readout process prepares a definite qubit state, we define the preparation fidelity $F_P$ as the conditional probability of $s_n = \varsigma$ given the estimator $\varsigma$ for the posterior qubit state $s_n$, i.e. $F_P^{\downarrow(\uparrow)} = P(s_n = \varsigma | \varsigma = \downarrow (\uparrow))$. We emulate the most relevant situation of a completely unknown input[8] by using data with 0.08 µs $< t_b < 1.3$ µs, for which $p_0^\downarrow = 0.500$. To optimally determine $\varsigma$ from $\mathbf{m}_n$, we again apply the Bayes' rule (Methods) and compare the likelihoods $P(\mathbf{m}_n | s_n = \downarrow)$ and $P(\mathbf{m}_n | s_n = \uparrow)$. We estimate $s_n$ from another estimator $\sigma'$ and convert the conditional probability $P(\sigma' = \varsigma | \varsigma = \downarrow (\uparrow))$ to $F_P^{\downarrow(\uparrow)}$ using the measurement fidelity of $\sigma'$ for $s_n$ (Methods). We obtain $F_P^{\downarrow(\uparrow)} =$ 76% (83%) for $n = 1$, which increments to 95.9% (88.6%) for $n = 20$ (Fig. 3d).

It is worth noting that for $n \geq 2$ these likelihoods $P(\mathbf{m}_n | s_n = \varsigma)$ can signal events where we have higher confidence in the final spin state. To explore this potential of heralded high-fidelity state preparation, we calculate the likelihood ratio $\Lambda^\varsigma = P(\mathbf{m}_{10} | s_{10} = \varsigma) / P(\mathbf{m}_{10} | s_{10} = \bar{\varsigma})$ (i.e. for $n = 10$) and select events with $\Lambda^\varsigma$ above a certain threshold. The conditional probability $P(\sigma' = \varsigma | \varsigma = \downarrow (\uparrow))$ is then estimated following the procedure described above (but with more ancilla measurements, see Methods). Indeed, $F_P^\downarrow$ increases from 94% to 99% at the median (for $\Lambda^\downarrow > 1$), and $F_P^\downarrow$ reaches 99.6% at the 76th percentile, see Fig. 4. The limiting value is higher for the spin-down case, as expected from $F_{QND}^{\downarrow(\uparrow)}$.

In the present experiment, 30 ancilla measurements are feasible before we lose strong correlation between the input and the outcome ($F_{QND}^\uparrow \lesssim 50\%$). This is limited by a relatively short electron spin lifetime, compared to single nuclear spins in silicon where 99.8% readout fidelity is achieved[18,19]. The ratio $T_1^\downarrow / T_1^\uparrow = 31$ is deviated from the ideal thermal population ratio ($= 16$) between the Zeeman sublevels at the electron temperature (~50 mK), and the measured $T_1^\uparrow$ is roughly 30 times shorter than nominal expectation for an idle spin away from the hotspot[20]. Indeed, data imply that the qubit relaxation occurs predominantly during the ancilla readout process (Supplementary Material). This effect is expected to be suppressed by further quenching the residual exchange coupling (~MHz) e.g. via an interdot gate electrode[6] or by fast readout with an ancilla encoded in double-dot spin states[21]. We anticipate that we will then improve $F_{QND}$ and the QND readout in all aspects, as a higher $F_{QND}$ should raise $F_M$ and $F_P$ that are achievable by repeating QND measurements.

$F_M$ and $F_P$ will also improve, particularly for small numbers of $n$, by decreasing single-shot QND measurement infidelities $1 - f_i^{\downarrow(\uparrow)}$, which are 15% (25%) on average for $i$ = 1-20. We estimate the contribution of charge discrimination error to be 7% for the spin-up case (Supplementary Material), which can be straightforwardly reduced by tuning the charge sensor sensitivity solely for the ancilla dot. Other errors in the qubit-ancilla entangling operation and the ancilla spin-to-charge conversion can be addressed by optimizing the two-qubit gate operation and the spin-selective tunneling process[4,7].

The ancilla-based QND readout is a crucial element in qubit error detection and correction protocols. Combined with high-fidelity single- and two-qubit gates[2,4], the demonstrated results will pave the way towards fault-tolerant quantum-information processing in the silicon quantum-dot platform.

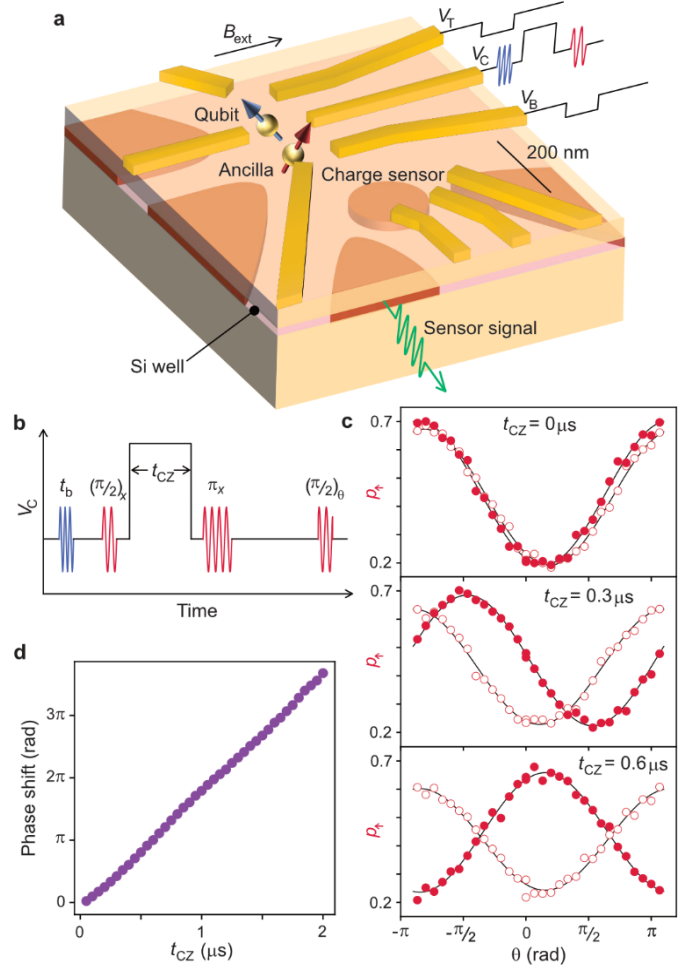

**Figure 1. Qubit and ancilla system | a,** Schematic of a device. The qubit spin (blue) and the ancilla spin (red) are hosted in two singly-occupied dots in a silicon quantum well layer. A proximal single electron transistor serves as a charge sensor. **b,** Control pulse. Two microwave tones (represented by different colors) are used to selectively rotate qubit and ancilla spins. A controlled-phase shift is induced by applying square pulses simultaneously to $V_T$, $V_B$ as well as to $V_C$. **c,** Ancilla spin-up probability after an entangling gate pulse. Traces with (without) a $\pi$ pulse applied to the qubit spin are plotted with filled (open) symbols. **d,** Measured controlled phase accumulation.

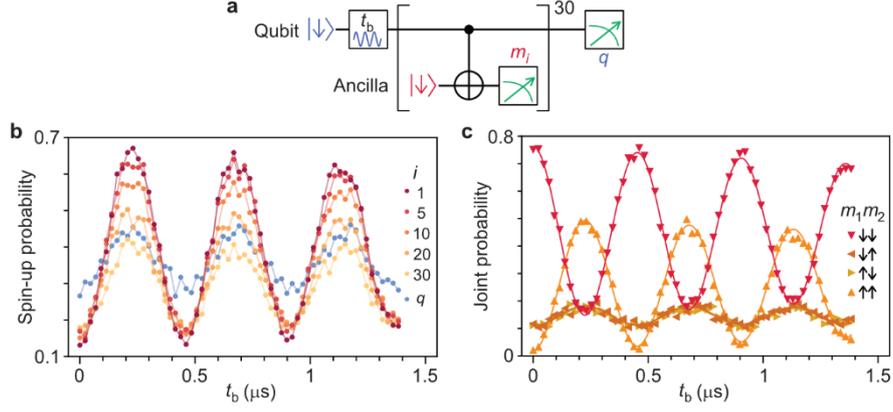

**Figure 2. Repetitive readout | a,** Quantum circuit for repetitive measurements. **b,** Spin-up probabilities of the *i*-th ancilla measurement (only $i = 1, 5, 10, 20$ and $30$ are shown for brevity) and the final qubit readout ($q$) out of 1000 events. Note the oscillation visibility for $q$ is influenced by the compromised sensor sensitivity. **c,** Probabilities of the four joint outcomes for the first and second ancilla measurements. The triangle symbols represent the experimental data, and the solid lines are the fit results to the model which takes into account preparation and measurement imperfections.

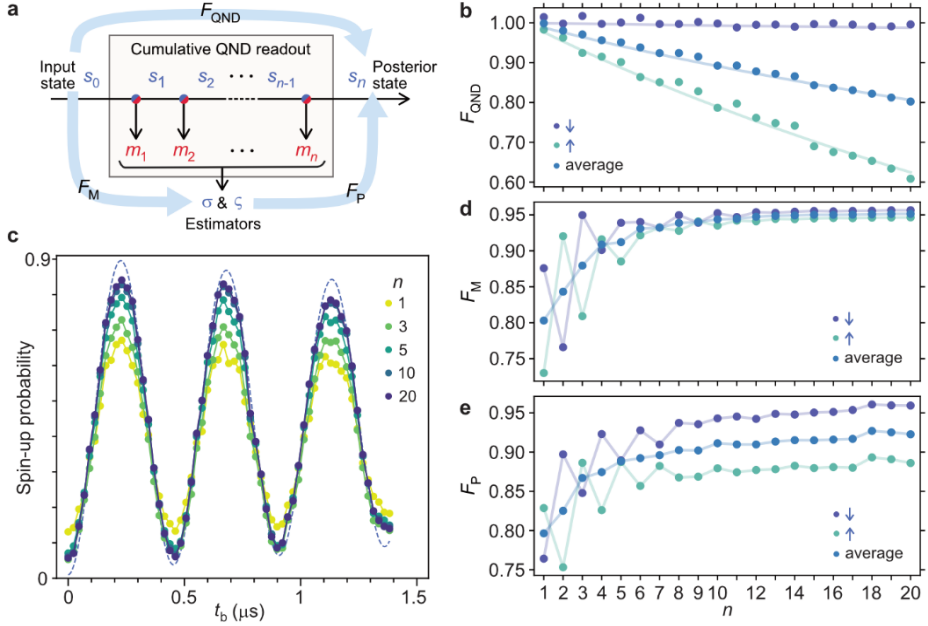

**Figure 3. Cumulative QND readout and fidelities | a,** Diagram for the cumulative readout protocol and fidelity definitions. We regard $n$ consecutive ancilla measurements (with outcomes $m_1, m_2 \ldots m_n$) as a single QND readout (with estimators $\sigma$ and $\varsigma$). The projected input spin state $s_0$ (either ↓ or ↑) changes to the posterior state $s_n$ after the process. The ideal QND measurement would give i) $s_n$ identical to $s_0$ (non-demolition), ii) $\sigma$ identical to $s_0$ (measurement) and iii) $s_n$ identical to $\varsigma$ (preparation). $F_{QND}$, $F_M$ and $F_P$ quantify these properties. **b,** $F_{QND}^{\downarrow}$, $F_{QND}^{\uparrow}$ and $(F_{QND}^{\downarrow} + F_{QND}^{\uparrow})/2$ after $n$ repetitive measurements. The solid lines show the values expected from the extracted $T_1^{\downarrow(\uparrow)}$. **c,** Rabi oscillations of the qubit spin acquired from multiple ancilla measurements. Plotted with a dashed curve is the estimated true qubit spin-up probability, $p_0^{\uparrow}$, consistent with a Rabi oscillation at a 630 kHz frequency detuning. **d,** $F_M^{\downarrow}$, $F_M^{\uparrow}$ and $(F_M^{\downarrow} + F_M^{\uparrow})/2$ as a function of $n$. State-dependent single-shot measurement fidelities $(f_i^{\downarrow} > f_i^{\uparrow})$ produce even-odd effects of $F_M^{\downarrow(\uparrow)}$, whereas the average increases monotonically. **e,** $F_P^{\downarrow}$, $F_P^{\uparrow}$ and $(F_P^{\downarrow} + F_P^{\uparrow})/2$ as a function of $n$.

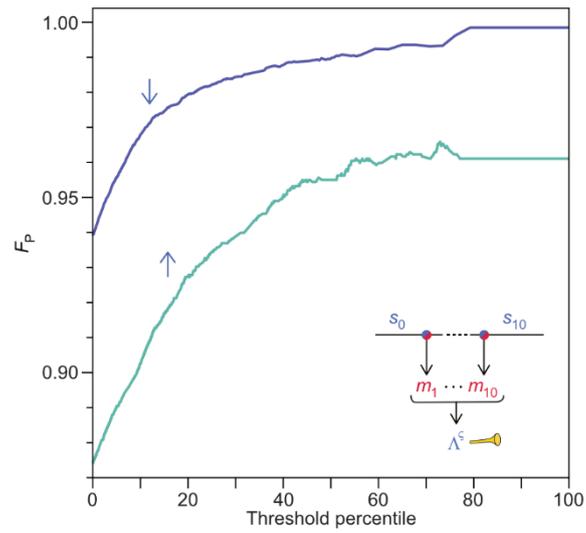

**Figure 4. Heralded enhanced preparation fidelity |** Events with high initialization confidence are selected based on $\Lambda^\varsigma$. The data include 53000 events in total and threshold percentiles with more than 5000 selected events are used for the analysis.

**References**

1. Veldhorst, M. *et al*., An addressable quantum dot qubit with fault-tolerant control-fidelity. *Nat. Nanotechnol.* **9**, 981-985 (2014).

2. Yoneda, J. *et al*., A quantum-dot spin qubit with coherence limited by charge noise and fidelity higher than 99.9%. *Nat. Nanotechnol.* **13**, 102-107 (2018).

3. Veldhorst, M. *et al*, A two-qubit logic gate in silicon. *Nature* **526**, 410-414 (2015).

4. Huang, W. *et al*., Fidelity benchmarks for two-qubit gates in silicon. *Nature* **569**, 532-536 (2019).

5. Watson, T. F. *et al*., A programmable two-qubit quantum processor in silicon. *Nature* **555**, 633-637 (2018).

6. Zajac, D.M., Hazard, T. M., Mi, X., Nielsen, E., Petta, J. R., Scalable Gate Architecture for a One-Dimensional Array of Semiconductor Spin Qubits. *Phys. Rev. Appl.* **6**, 054013 (2016).

7. Keith, D. *et al*., Benchmarking high fidelity single-shot readout of semiconductor qubits. arXiv: 1811.03630.

8. Ralph, T. C., Bartlett, S. D., O'Brien, J. L., Pryde, G. J., Wiseman, H. M., Quantum nondemolition measurements for quantum information. *Phys. Rev.* A**73**, 012113 (2006).

9. Fowler, A. G., Mariantoni, M., Martinis, J. M., Cleland, A. N., Surface codes: Towards practical large-scale quantum computation. *Phys. Rev.* A **86**, 032324 (2012).

10. Devoret, M. H., Schoelkopf, R. J., Superconducting Circuits for Quantum Information: An Outlook. *Science* **339**, 1169-1175 (2013).

11. Takeda, K. *et al*., A fault-tolerant addressable spin qubit in a natural silicon quantum dot. *Sci. Adv.* **2**, e1600694 (2016).

12. Yoneda, J. *et al*., Robust micromagnet design for fast electrical manipulations of single spins in quantum dots. *Appl. Phys. Express* **8**, 84401 (2015).

13. Meunier, T., Calado, V. E., Vandersypen, L. M. K., Efficient controlled-phase gate for single-spin qubits in quantum dots. *Phys. Rev.* B **83**, 121403 (2011).

14. Reed, M. D. *et al*., Reduced Sensitivity to Charge Noise in Semiconductor Spin Qubits via Symmetric Operation. *Phys. Rev. Lett.* **116**, 110402 (2016).

15. Yoneda, J. *et al.*, Fast electrical control of single electron spins in Quantum dots with vanishing influence from nuclear spins. *Phys. Rev. Lett.* **113**, 267601 (2014).

16. Takeda, K. *et al*., Optimized electrical control of a Si/SiGe spin qubit in the presence of an induced frequency shift. *npj Quantum Inf.* **4**, 54 (2018).

17. Nakajima, T. *et al.*, Quantum non-demolition measurement of an electron spin qubit. *Nature Nanotechnol.* **14**, 555–560 (2019).

18. Pla, J. J. *et al.* High-fidelity readout and control of a nuclear spin qubit in silicon. *Nature* **496**, 334-338 (2013).

19. Hensen, B. *et al.*, A silicon quantum-dot-coupled nuclear spin qubit. arXiv:1904.08260.

20. Borjans, F., Zajac, D. M., Hazard, T. M., Petta, J. R., Single-spin relaxation in a synthetic spin-orbit field. *Phys. Rev. Appl.* **11**, 044063 (2019).

21. Noiri, A. *et al.*, A fast quantum interface between different spin qubit encodings. *Nat. Commun.* **9**, 5066 (2018).

**Acknowledgements** We thank Microwave Research Group at Caltech for technical assistance. Part of this work was financially supported by CREST, JST (JPMJCR15N2, JPMJCR1675), the ImPACT Program of Council for Science, Technology and Innovation (Cabinet Office, Government of Japan), Q-LEAP project initiated by MEXT, Japan, JSPS KAKENHI Grants Nos. 26220710, 17K14078, 18H01819, and 19K14640, RIKEN Incentive Research Projects and The Murata Science Foundation.

**Author contributions** J.Y. acquired and analyzed the data and wrote the manuscript. K.T. and A.N. set up the measurement hardware. T.N. contributed to the data analysis. K.T. fabricated the device with help from J.K. and T.K. All authors discussed the result. S.T. supervised the project.